\documentclass{PoS}

\title{Nuclear forces from quenched and 2+1 flavor lattice QCD
using the PACS-CS gauge configurations}

\ShortTitle{Nuclear forces from quenched and 2+1 flavor lattice QCD}

\author{\speaker{Noriyoshi Ishii}\\
  Center for Computational Sciences, University of Tsukuba,\\
  Tsukuba, Ibaraki 305--8577, Japan\\
  E-mail: \email{ishii@ribf.riken.jp}}
\author{Sinya Aoki\\
  Graduate School of Pure and Applied Sciences, University of Tsukuba,\\
  Tsukuba, Ibaraki 305--8577, Japan\\
  RIKEN BNL Research Center, Brookhaven National Laboratory,\\
  Upton, New York 11973, USA\\
  E-mail: \email{saoki@het.ph.tsukuba.ac.jp}}
\author{Tetsuo Hatsuda\\
  Department of Physics, University of Tokyo,\\
  Tokyo 113--0033, Japan\\
  E-mail: \email{hatsuda@phys.s.u-tokyo.ac.jp}}
\author{for PACS-CS Collaboration}

\abstract{
Two of  recent progress in lattice  QCD approach to  nuclear force are
reported.
(i)  Tensor  force  from  quenched  lattice  QCD:  By  truncating  the
derivative expansion of inter-nucleon  potential to the strictly local
terms, we obtain central force $V_{\rm C}(r)$ and tensor force $V_{\rm
T}(r)$ separately from s-wave  and d-wave components of Bethe-Salpeter
wave function for two nucleon state with $J^P=1^+$.
Numerical calculation is performed with quenched QCD on $32^4$ lattice
using the  standard plaquette action at $\beta=5.7$  with the standard
Wilson quark action with $\kappa=0.1640, 0.1665, 0.1678$.
Preliminary results show that the depths of the resulting tensor force
amount to  20 to  40 MeV, which  is enhanced  in the light  quark mass
region.
(ii)   Nuclear  force  from   2+1  flavor   QCD  with   PACS-CS  gauge
configuration: Preliminary full QCD  results are obtained by using 2+1
flavor gauge configurations generated by PACS-CS collaboration.
The resulting  potential has the  midium range attraction of  about 30
MeV  similar to  the  preceding quenched  calculations.  However,  the
repulsive core  at short distance  is significantly stronger  than the
corresponding quenched QCD result.  }

\FullConference{The XXVI International Symposium on Lattice Field Theory\\
		 July 14-19 2008\\
		 Williamsburg, Virginia, USA}

\newcommand{\Sect}[1]{Sect.~\ref{#1}}
\newcommand{\Eq}[1]{Eq.~(\ref{#1})}

\newcommand{\Fig}[1]{Fig.~\ref{#1}}

\newcommand{\alt}{\;\raisebox{0.5ex}{\mbox{$<$}}\hspace*{-0.7em} \raisebox{-0.5ex}{\mbox{$\sim$}}\;}
\newcommand{\agt}{\;\raisebox{0.5ex}{\mbox{$>$}}\hspace*{-0.7em} \raisebox{-0.5ex}{\mbox{$\sim$}}\;}
\begin{document}

\section{Introduction}
Proton and neutron are the fundamental constituents of atomic nucleus.
The interaction among them are  called as the nuclear force.
Together with the  structures of nuclei, the nuclear  force itself has
been  actively investigated  in  nuclear physics  since its  discovery
about 75 yeas ago.
Phenomenologically,   the  properties   of  the   nuclear   force  are
characterized by the attraction at  medium distance with the OPEP tail
\cite{yukawa,taketani} and the strong repulsive core \cite{jastrow} at
short distance.
The former  is responsible for nuclei  to be bound.
The latter is  important for various phenomena, such  as the stability
of atomic  nuclei, super nova explosions  of type II,  and the maximum
mass of neutron star.
While the medium  to long distance properties are  accessible with the
meson exchange picture, it is difficult to approach the short distance
properties. In  particular, the origin  of the repulsive core  has not
yet been answered beyond phenomenological models.
Since the nucleons  overlaps at short distance, the  repulsive core is
expected to reflect the  internal structure of nucleon. Therefore, QCD
is  considered  to be  the  best tool  to  reveal  the short  distance
properties of the  nuclear force.
Indeed, one attempted to study  the repulsive core with lattice QCD by
extending the method of static quark potential \cite{takahashi}.
However, the  repulsive core was  not reproduced from  this pioneering
work.

Recently, we have developed a  new method to extract the nuclear force
between  nucleons composed of  non-static quarks  and have  found that
essential features of the nuclear force are reproduced \cite{ishii}.
In this  method, lattice  QCD is used  to generate  the Bethe-Salpeter
(BS)  wave function  for a  two nucleon  state in  the center  of mass
frame:
\begin{equation}
  \psi_{\alpha\beta}(\vec x - \vec y)
  \equiv
  \lim_{t\to +0}
  \left\langle 0 \left|
  T\left[
    p_{\alpha}(\vec x, t)
    n_{\beta} (\vec y, 0)
    \right]
  \right| NN \right\rangle,
  \label{bs.amplitude}
\end{equation}
where    $p_{\alpha}\equiv   \epsilon_{abc}\left(    u_a^T   C\gamma_5
d_b\right)  u_{c;\alpha}$  and  $n_{\beta}\equiv  \epsilon_{abc}\left(
u_a^T  C\gamma_5 d_b\right)  d_{c;\beta}$ denote  interpolating fields
for  proton and neutron,  respectively. Note  that this  represents an
amplitude to find  three quarks at $\vec x$ and  other three quarks at
$\vec y$.
At large  separation, i.e.,  $|\vec x -  \vec y|\to  \mbox{large}$, it
shows a  desirable asymptotic behavior, which is  characterized by the
scattering phase shift $\delta(k)$ as
\begin{equation}
  \psi(\vec r)
  \sim
  \frac{\sin(kr + \delta(k))}{kr}
  + \cdots
  \hspace{0.1\textwidth} \mbox{for s-wave}.
  \label{asymptotic.form}
\end{equation}
Here, $k$  corresponds to  the ``asymptotic momentum''  measured beyond
the range  of the  interaction, which is  related to the  total energy
$P_0$ of the state in \Eq{bs.amplitude} as $P_0 = 2\sqrt{m_{\rm N}^2 +
k^2}$.
The amplitude \Eq{bs.amplitude}  satisfies the effective Schr\"odinger
equation as
\begin{equation}
  (\vec \nabla^2 + k^2) \psi(\vec r)
  =
  m_{\rm N}
  \int d^3 r'
  V_{NN}(\vec r, \vec r')
  \psi(\vec r').
  \label{effective.schrodinger.eq}
\end{equation}
(For  derivation, see  Ref.~\cite{csd}.) In  the r.h.s.,  $V_{\rm NN}$
plays  the role  of  the  interaction kernel.   It  is most  generally
non-local, and can be defined to be independent of the total energy of
the  state  \cite{csd}.
%
After  the  constraints  from   various  symmetris  are  imposed,  the
derivative expansion leads us to
\begin{equation}
  V_{\rm NN}(\vec r,\vec r')
  =
  \left\{
  V_{\rm C}(r)
  + V_{\rm T}(r) S_{12}
  + V_{\rm LS}(r) \vec L\cdot \vec S
  + O(\nabla^2)
  \right\}
  \delta(\vec r - \vec r').
  \label{derivative.expansion}
\end{equation}
Here,   $S_{12}\equiv   3(\vec   \sigma_1\cdot\vec  r)(\vec   \sigma_2
\cdot\vec r)/r^2 - \vec  \sigma_1\cdot\vec\sigma_2$, $\vec L \equiv -i
\vec r  \times \vec \nabla$, and  $\vec S\equiv (\vec  \sigma_1 + \vec
\sigma_2)/2$.
$V_{\rm C}(r)$, $V_{\rm T}(r)$ and  $V_{\rm LS}(r)$ are referred to as
``central force'',  ``tensor force'',  and ``LS force''.   
(Iso-spin  dependence  of  these   potentials  are  understood  to  be
implicit.)
These three  forces play the  most important role in  the conventional
nuclear physics.
By  truncating \Eq{derivative.expansion}  up  to the  first term,  the
central  potential $V_{\rm  C}(r;  ^1S_0)$ and  the effective  central
potential $V_{\rm  C}^{\rm eff}(r;  ^3S_1)$ have been  calculated from
quenched lattice  QCD \cite{ishii}.  The  resulting potentials possess
the  repulsive core at  short distance  as well  as the  attraction at
medium distance \cite{ishii}.
Note that,  owing to \Eq{asymptotic.form}, the method  can be extended
to be more faithful to  NN scattering experiments \cite{csd}.

The  contents are  organized  as follows.
In \Sect{section.two}, we extend our  method to tensor force, which is
obtained  from Schr\"odinger  equation in  coupled $^3  S-D_1$ partial
waves in $J^P=1^+$.
In  \Sect{section.three}, we give  2+1 flavor  lattice QCD  results of
nuclear force by using PACS-CS gauge configurations.
%

\section{Tensor force from quenched lattice QCD}
\label{section.two}
Tensor force plays an important role in nuclear physics. Together with
the repulsive core, it has  important influences on the structures and
the stabilities of nuclei.  However, phenomenological determination of
tensor force is  known to be afflicted with  an uncertainty especially
at short distance due to the existence of centrifugal barrier.

To  obtain tensor  force  in lattice  QCD,  we consider  Schr\"odinger
equation for $J^P=1^+$.
In  this case,  the wave  function  has two  components, i.e.,  s-wave
component and d-wave component.  The central force $V_{\rm C}(r)$ acts
separately  within these  two  components.  The  tensor force  $V_{\rm
T}(r)$ provides a coupling between these two, and the action of the LS
force $V_{\rm LS}(r)$ is restricted within the d-wave component.
If we  keep only the  first term in \Eq{derivative.expansion},  we can
obtain  only  the  effective  central force  $V_{\rm  C}^{\rm  eff}(r;
^3S_1)$.
By  keeping the  one  more term  in  \Eq{derivative.expansion}, it  is
possible to obtain the central force $V_{\rm C}(r)(\equiv V_{\rm C}(r;
^3S_1))$ and  the tensor force  $V_{\rm T}(r)$ separately.   Note that
these    two   terms    give   strictly    local    contributions   in
\Eq{derivative.expansion}.
The effective Schr\"odinger equation \Eq{effective.schrodinger.eq} becomes
\begin{equation}
  (H_0 + V_{\rm C}(r) + V_{\rm T}(r) S_{12})
  \psi(\vec r)
  =
  E
  \psi(\vec r),
  \label{schrodinger.eq.one.plus}
\end{equation}
where $H_0\equiv  - \nabla^2/m_{\rm  N}$, and $E\equiv  k^2/m_{\rm N}$
denotes  the non-relativistic energy.   For definiteness,  we restrict
ourselves to  BS wave  function $\psi(\vec r)$  for an state  with the
azimuthal quantum number $M=0$, i.e.,
\begin{equation}
  \psi_{\alpha\beta}(\vec x - \vec y)
  \equiv
  \lim_{t\to +0}
  \left\langle 0 \left|
  T\left[
    p_{\alpha}(\vec x, t)
    n_{\beta}(\vec y,  0)
    \right]
  \right|NN(J^P=1^+; M=0)\right\rangle.
\end{equation}
We define projection  operators ${\cal P}$ and ${\cal  Q}$ onto s-wave
and d-wave components, respectively, as
\begin{equation}
  {\cal P}\psi_{\alpha\beta}(\vec r)
  \equiv
  \frac1{24}
  \sum_{g\in O}
  \psi_{\alpha\beta}(g^{-1}\vec r),
  \hspace*{0.1\textwidth}
  {\cal Q} \equiv 1 - {\cal P}.
\end{equation}
where $O$ denotes the cubic group, which consists of 24 elements.
We multiply ${\cal P}$  and ${\cal Q}$ to \Eq{schrodinger.eq.one.plus}
from the left. Since $H_0$,  $V_{\rm C}(r)$ and $V_{\rm T}(r)$ commute
with  ${\cal P}$ and  ${\cal Q}$,  \Eq{schrodinger.eq.one.plus} splits
into the following two equations as
\begin{eqnarray}
  H_0 [{\cal P} \psi](\vec r)
  + V_{\rm C}(r) [{\cal P}\psi](\vec r)
  + V_{\rm T}(r) [{\cal P} S_{12} \psi](\vec r)
  &=&
  E
  [{\cal P} \psi](\vec r)
  \\
  H_0 [{\cal Q} \psi](\vec r)
  + V_{\rm C}(r) [{\cal Q}\psi](\vec r)
  + V_{\rm T}(r) [{\cal Q} S_{12} \psi](\vec r)
  &=&
  E
  [{\cal Q} \psi](\vec r).
\end{eqnarray}
Note that each  of these equations has two  Dirac indices $\alpha$ and
$\beta$.
For  definiteness,  we pick  up  $(\alpha,\beta)=(1,0)$ components  of
these two  equations, and  solve them for  $V_{\rm C}(r)$  and $V_{\rm
T}(r)$. We arrive at
\begin{eqnarray}
  V_{\rm C}(\vec r)
  &=&
  E
  + \frac1{\Delta(\vec r)}
  \left(
  [{\cal Q} S_{12}\psi](\vec r) H_0 [{\cal P}\psi](\vec r)
  -
  [{\cal P} S_{12}\psi](\vec r) H_0 [{\cal Q}\psi](\vec r)
  \right)
  \label{eq.vc}
  \\\nonumber
  V_{\rm T}(\vec r)
  &=&
  \frac1{\Delta(\vec r)}
  \left(
  -
  [{\cal Q} \psi](\vec r) H_0 [{\cal P} \psi](\vec r)
  +
  [{\cal P} \psi](\vec r) H_0 [{\cal Q} \psi](\vec r)
  \right),
\end{eqnarray}
with $\Delta(\vec  r) \equiv [{\cal  P}\psi](\vec r) [{\cal  Q} S_{12}
\psi](\vec r)  - [{\cal Q}  \psi](\vec r) [{\cal P}  S_{12} \psi](\vec
r)$. Note that, if the d-wave component in the wave function vanishes,
the  first  line  in  \Eq{eq.vc}  reduces  to $V_{\rm  C}(r)  =  (E  -
H_0)\psi(\vec r)/\psi(\vec r)$.

Numerical  calculation is performed  with quenched  QCD by  using Blue
Gene/L  at KEK.  The  quenched gauge  configurations are  generated by
employing the  standard plaquette  gauge action at  $\beta=5.7$, which
leads     to     the     lattice     spacing     $a^{-1}=1.44$     GeV
\cite{fukugita}. Propagators of quarks  are generated by employing the
standard   Wilson   quark   action   with   the   hopping   parameters
$\kappa=0.1640,  0.1665, 0.1678$,  which correspond  to $m_{\pi}\simeq
731, 529, 380$ MeV,  respectively. These calculations are performed by
using $N_{\rm  conf}= 1000, 2000, 2021$ gauge  configurations. BS wave
functions   are   picked  up   from   the  time-slice   $t-t_0=9,8,6$,
respectively, where the ground state saturation is achieved within the
error bars. While the periodic boundary condition is imposed along the
spatial directions, Dirichlet boundary  condition is imposed along the
temporal direction  on the time-slice  $t=0$.  Wall source is  used on
the time-slice $t=t_0\equiv 5$ after imposing Coulomb gauge.

\begin{figure}[h]
\begin{center}
\includegraphics[height=0.48\textwidth,angle=-90]{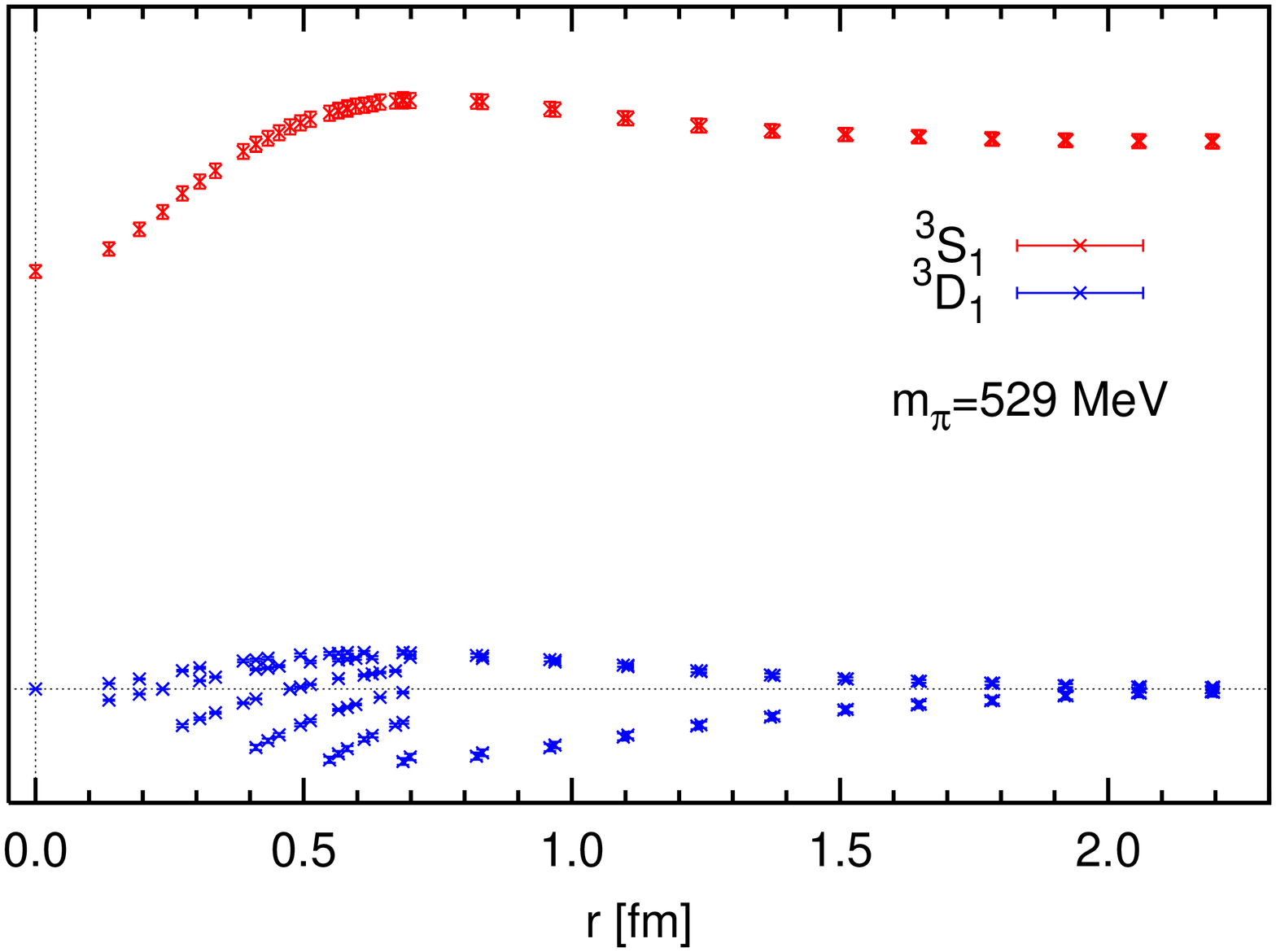}
\includegraphics[height=0.48\textwidth,angle=-90]{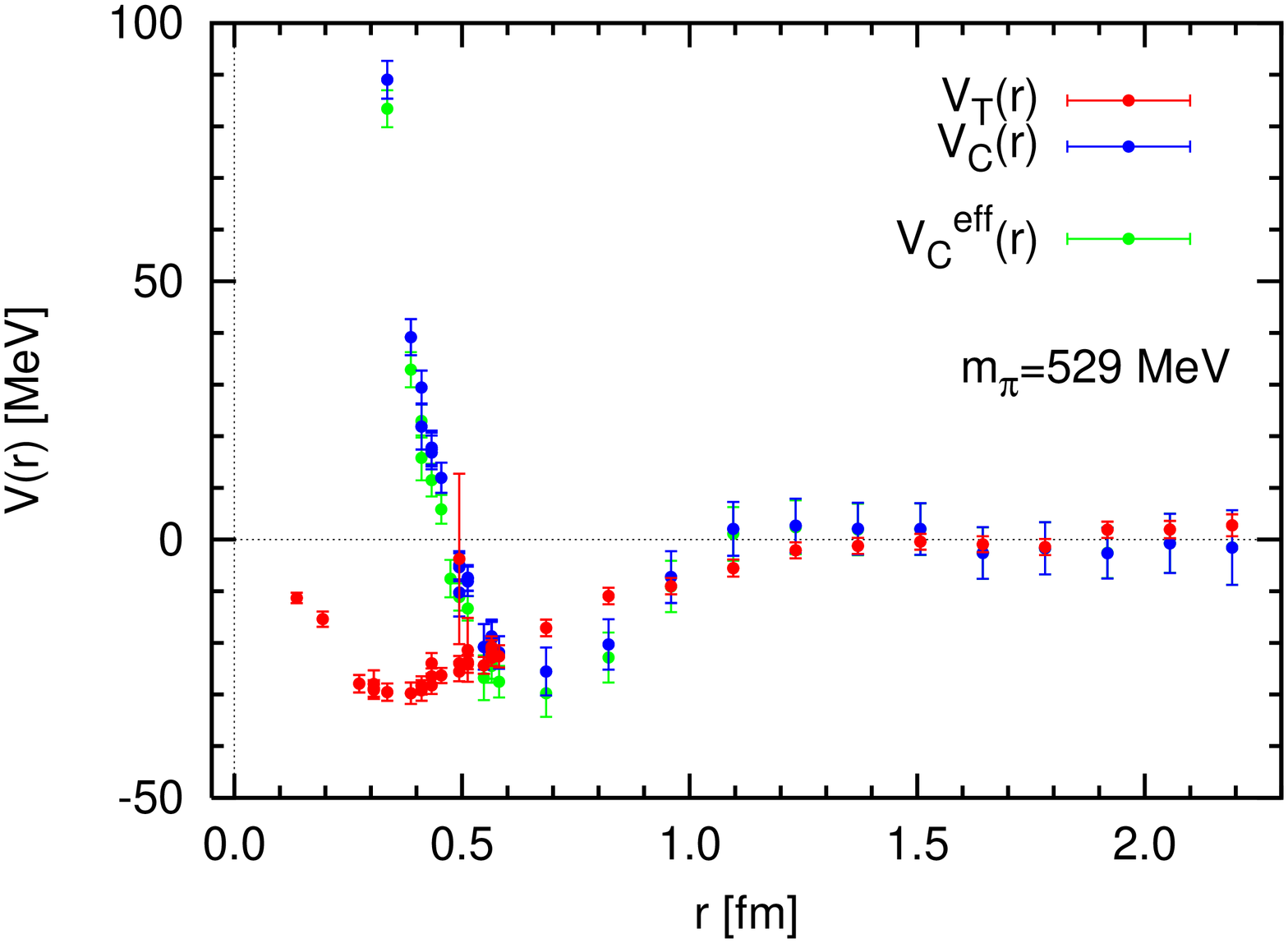}
\end{center}
\caption{$(\alpha,\beta)=(1,0)$  part  of the  s-wave  and the  d-wave
components of BS wave function  for a state with $J^P=1^+, M=0$ (left)
and  the  reconstructed inter-nucleon  potentials  $V_{\rm C}(r)$  and
$V_{\rm T}(r)$ (right). }
\label{fig.tensor.force}
\begin{center}
\includegraphics[height=0.48\textwidth,angle=-90]{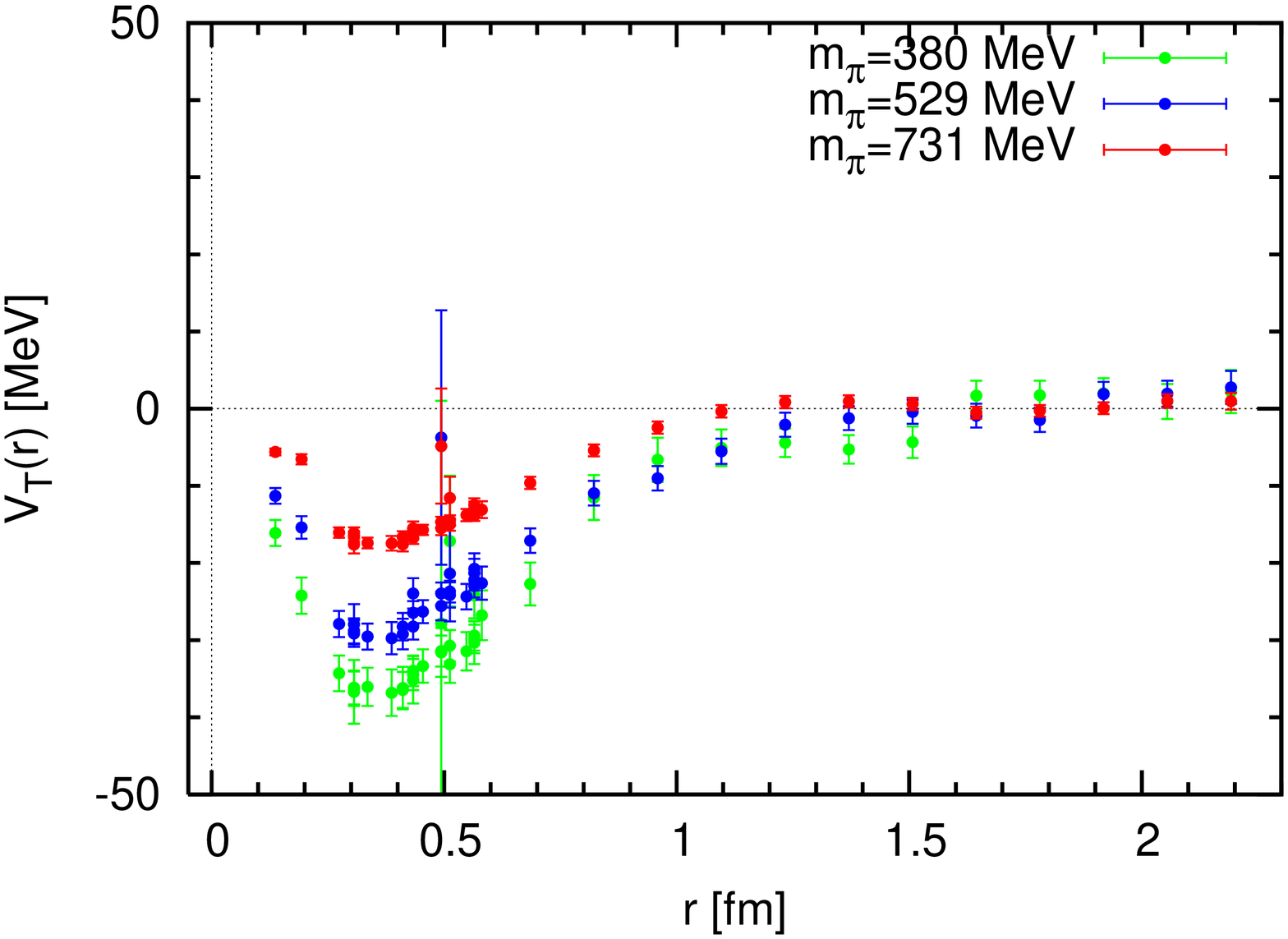}
\end{center}
\caption{Quark mass dependence of tensor force.}
\label{fig.tensor.force.two}
\end{figure}
The  l.h.s.  of \Fig{fig.tensor.force}  shows BS  wave functions  of a
$J^P=1^+,  M=0$ state  for  $m_{\pi}\simeq 529$  MeV.   To reduce  the
calculational  cost, calculation is  restricted to  the points  on the
coordinate  axes and  their nearest  neighbors  for $r  \agt 0.7$  fm,
whereas all points are calculated for $r \alt 0.7$ fm.
Note  that $^3D_1$  part  of the  wave  function is  multivalued as  a
function of $r$, which is due to the angular dependence.
The  r.h.s.   of \Fig{fig.tensor.force}  shows  the resulting  central
force  $V_{\rm C}(r)$ and  tensor force  $V_{\rm T}(r)$  together with
effective central force $V_{\rm C}^{\rm eff}(r) \equiv (E - H_0)[{\cal
P}\psi](\vec r)/[{\cal P}\psi](\vec r)$.
From viewpoint  of the meson  exchange picture, this shape  of $V_{\rm
T}(r)$ is expected from  a cancellation between the contributions from
pion and rho meson.
Note  that $V_{\rm  C}^{\rm eff}(r)$  contains the  effect  of $V_{\rm
T}(r)$ through the 2nd order perturbation,
where  one expects  that $V_{\rm  C}^{\rm eff}(r)$  acquires sufficient
attraction to generate the bound deuteron.
However, we see that the difference between $V_{\rm C}(r)$ and $V_{\rm
C}^{\rm eff}(r)$  is quite small, which  may be due to  an artifact of
heavy quark mass.
\Fig{fig.tensor.force.two} shows  the quark mass  dependence of tensor
force  $V_{\rm T}(r)$. We  see that  tensor force  is enhanced  in the
light  quark mass  region,  which suggests  the  importance of  direct
lattice QCD calculation employing light quark mass.

A technical comment is in order.  Recall that we used the spin $(1,0)$
component of  \Eq{eq.vc}, where the  2nd equation vanishes at  $\vec r
\propto  (\pm 1,\pm  1,\pm  1)$.   This is  because  the spin  $(1,0)$
component of  d-wave part in the  wave function for  $J^P=1^+, M=0$ is
proportional         to         the        spherical         harmonics
$Y_{M=0}^{(l=2)}(\theta,\phi)\propto   3\cos^2\theta   -   1$,   which
vanishes at $\vec r \propto (\pm 1,\pm 1,\pm 1)$.
Although these points are removed from the plots, statistical error is
accumulated in  the neighborhood of these points.   (For instance, see
the  points at  $r\simeq 0.5$  fm in  Figs.~\ref{fig.tensor.force} and
\ref{fig.tensor.force.two}.)  It  is desirable to improve  this in the
near future.

\section{Nuclear force from 2+1 flavor lattice QCD}
\label{section.three}
To compare our results with empirical  data, a key role is played by a
full QCD calculation on a large volume employing a smaller quark mass.
PACS-CS collaboration is generating 2+1 flavor gauge configurations on
a large volume in significantly light quark mass region \cite{pacscs}.
PACS-CS gauge configurations are  generated by employing Iwasaki gauge
action at  $\beta=1.90$ on  $32^3\times 64$ lattice  and O(a)-improved
Wilson  quark  (clover)  action  with  a  non-perturbatively  improved
coefficient $c_{\rm SW}=1.715$ \cite{pacscs}.
The  lattice   scale  is   determined  from  $m_{\pi}$,   $m_{K}$  and
$m_{\Omega}$ inputs leading to $a^{-1}=2.176(31)$ GeV ($a\simeq 0.091$
fm). Hence,  the spatial extension amounts  to $L = 32  a \simeq 2.90$
fm.
To  calculate  nuclear force,  we  use  two  series of  PACS-CS  gauge
configurations        with        $(\kappa_{\rm        ud},\kappa_{\rm
s})=(0.13700,0.13640)$  and $(0.13770,0.13640)$,  which  correspond to
$m_{\pi} \simeq 702, 296$ MeV, respectively.

To  calculate  BS  wave  function,  we impose  the  periodic  boundary
condition along the  spatial direction.  On the other  hand, along the
temporal direction, we impose  the Dirichlet boundary condition on the
time-slice $t=32(=N_t/2)$. We locate the wall source on the time-slice
$t=0$ with Coulomb gauge.
Note that the setup is strictly symmetric around the hyper-plane $t=0$
aiming at doubling the number  of data by using the charge conjugation
and the time-reversal. (See below.)

We consider the Euclidean four point correlator of nucleon fields with
wall source as
\begin{equation}
  G_{\alpha\beta;\alpha'\beta'}(\vec x,\vec y,t)
  \equiv
  \left\langle 0 \left|
  T\left[
    p_{\alpha}(\vec x,t)
    n_{\beta}(\vec y, t)
    {\bar p}'_{\alpha'}
    {\bar n}'_{\beta'}
    \right]
  \right| 0 \right\rangle,
\end{equation}
where   ${\bar  p}'_{\alpha}\equiv  \sum_{\vec   x,\vec  y,   \vec  z}
\epsilon_{abc}\left( {\bar u}_a(\vec x) C\gamma_5 {\bar d}^T_b(\vec y)
\right)  {\bar u}_{c;\alpha}(\vec  z)$ and  ${\bar  n}'_{\beta} \equiv
\sum_{\vec  x,\vec y,\vec z}  \epsilon_{abc}\left( {\bar  u}_a(\vec x)
C\gamma_5  {\bar d}^T_b(\vec  y)\right) {\bar  d}_{d;  \beta}(\vec z)$
denote wall sources for proton and neutron, respectively.
%
%
Statistical  noises  are  reduced  by  utilizing  the  following  four
symmetries.
(i)
  The spatial  translation:
$
    G_{\alpha\beta;\alpha'\beta'}(\vec x,\vec y, t)
$
$
    =
$
$
    G_{\alpha\beta;\alpha'\beta'}(\vec x+\vec\Delta, \vec y + \vec \Delta, t),
$
  where $\vec \Delta$ denotes an arbitrary 3 dimensional vector.
(ii)
  The cubic group:
$
    G_{\alpha\beta;\alpha'\beta'}(\vec x,\vec y, t)
$
$
    =
$
$
    S_{\alpha\tilde\alpha}(g)
$
$
    S_{\beta\tilde\beta}(g)
$
$
    G_{\tilde\alpha\tilde\beta;\tilde\alpha'\tilde\beta'}
    (g^{-1}\vec x,g^{-1}\vec y, t)
$
$
    S_{\tilde\alpha'\alpha'}(g^{-1})
$
$
    S_{\tilde\beta'\beta'}(g^{-1}),
$
where $g$ denotes an arbitrary element of the cubic group.
$S(g)$ denotes  the (double-valued) representation matrix  of SO(3) in
the  Dirac  bispinor  space, i.e.,  $S(g)\equiv  \exp\left(\frac{i}{4}
\sigma_{ij}\omega_{ij}\right)$  with $\sigma_{ij}  \equiv -\frac{i}{2}
[\gamma_i,\gamma_j]$   for   $g    =   e^{\omega}$   with   $\omega\in
\mbox{so}(3)$.
(iii)
  The spatial  reflection:
$
    G_{\alpha\beta;\alpha'\beta'}(\vec x,\vec y, t)
$
$
    =
$
$
    (\gamma_0)_{\alpha\tilde\alpha}
$
$
    (\gamma_0)_{\beta \tilde\beta}
$
$
    G_{\tilde \alpha \tilde\beta; \tilde\alpha' \tilde\beta'}
    (-\vec x,-\vec y, t)
$
$
    (\gamma_0)_{\tilde\alpha' \alpha'}
$
$
    (\gamma_0)_{\tilde\beta'  \beta'}.
$
(iv)
  The charge conjugation  and time-reversal:
$
    G_{\alpha\beta;\alpha'\beta'}(\vec x,\vec y, t)
$
$
    =
$
$
    (-C\gamma_0)_{\alpha\tilde\alpha}
$
$
    (-C\gamma_0)_{\beta\tilde\beta}
$
$
    G^{*}_{\tilde\alpha\tilde\beta;\tilde\alpha'\tilde\beta'}
    (\vec x,\vec y, -t)
$
$
    (-C\gamma_0)_{\tilde\alpha'\alpha'}
$
$
    (-C\gamma_0)_{\tilde\beta'\beta'}.
$
  Note that QCD Lagrangian has the charge conjugation symmetry: ${\cal
  C}q{\cal  C}^{-1}  \equiv   C{\bar  q}^T$,  ${\cal  C}{\bar  q}{\cal
  C}^{-1}\equiv  q^T C$,  ${\cal  C} A_{\mu}  {\cal  C}^{-1} \equiv  -
  A_{\mu}^T$, where the matrix notation  of the gluon field is adopted
  as $A_{\mu}  \equiv A_{\mu}^a T^a$ for color  SU(3) generator $T^a$.
  This implies the charge  conjugation of the composite nucleon fields
  ${\cal C}  N {\cal C}^{-1} \equiv  -C \bar N^T$ and  ${\cal C}\bar N
  {\cal  C}^{-1} =  - N^T  C$, where  $N$ represents  an interpolating
  field  for  proton  or  neutron.   Note  that,  for  correlators  in
  imaginary  time,  the  charge   conjugation  is  combined  with  the
  time-reversal through the complex conjugation.

The BS wave function $\psi_{\alpha\beta}(\vec r)$ for the ground state
is obtained from the Euclidean  four point correlator in the large $t$
region after multiplied by ${\bf P}_{\alpha'\beta'}(J^P,M)$ as
\begin{equation}
  G_{\alpha\beta;\alpha'\beta'}(\vec x,\vec y,t)
  {\bf P}_{\alpha'\beta'}(J^P,M)
  =
  A_0 \psi_{\alpha\beta}(\vec x - \vec y; J^P,M) e^{-E_0 t}
  +
  \cdots,
\end{equation}
where          ${\bf          P}_{\alpha'\beta'}(J^P,M)         \equiv
(\sigma_2)_{\alpha'\beta'}$  and  $(\sigma_2\sigma_M)_{\alpha'\beta'}$
for  $J^P=0^+$  and   $1^+$,  respectively.   ``$\cdots$''  represents
contributions from excited  states, which are exponentially suppressed
in the large $t$ region.

\begin{figure}[h]
\begin{center}
\includegraphics[height=0.48\textwidth,angle=-90]{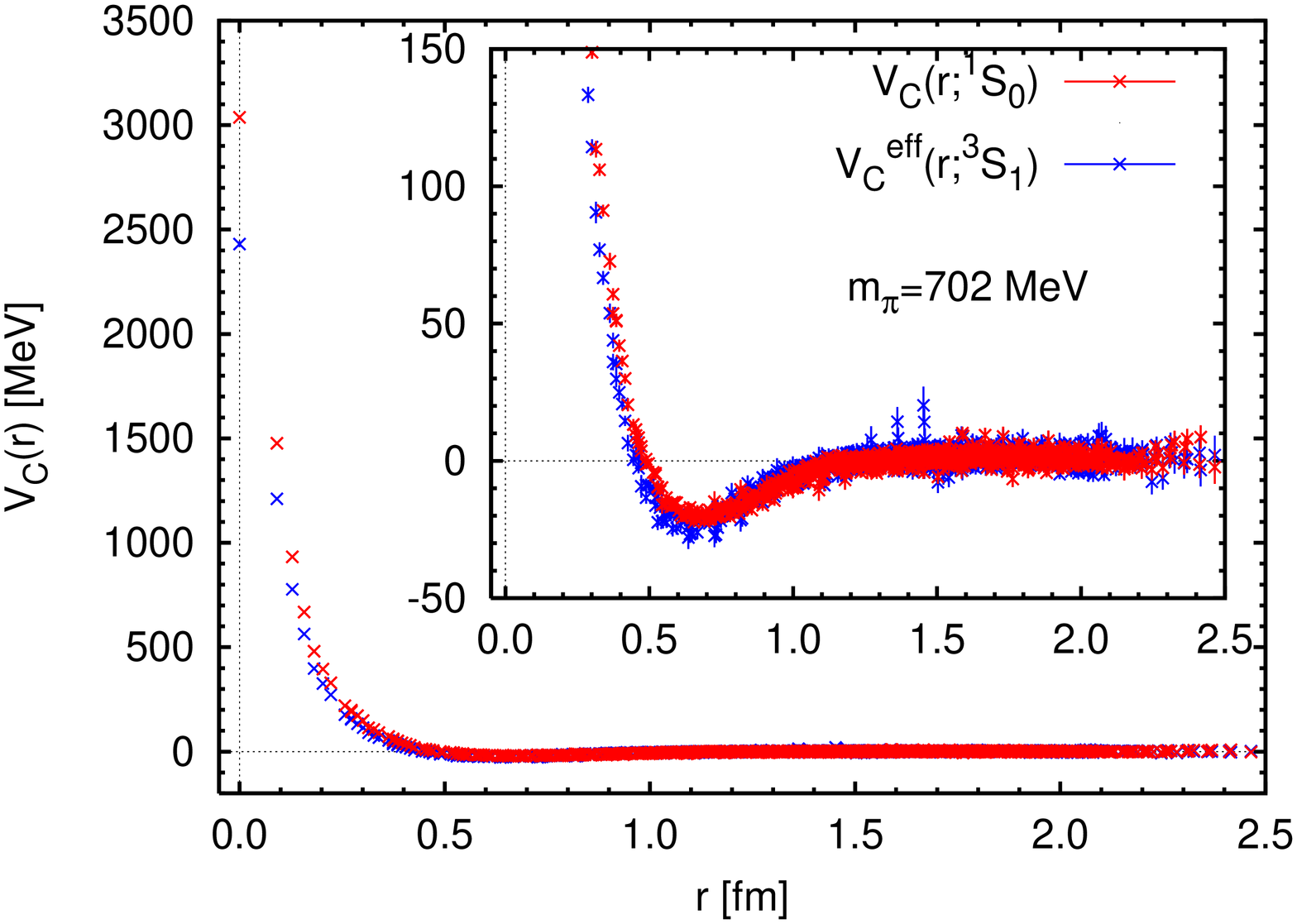}
\includegraphics[height=0.48\textwidth,angle=-90]{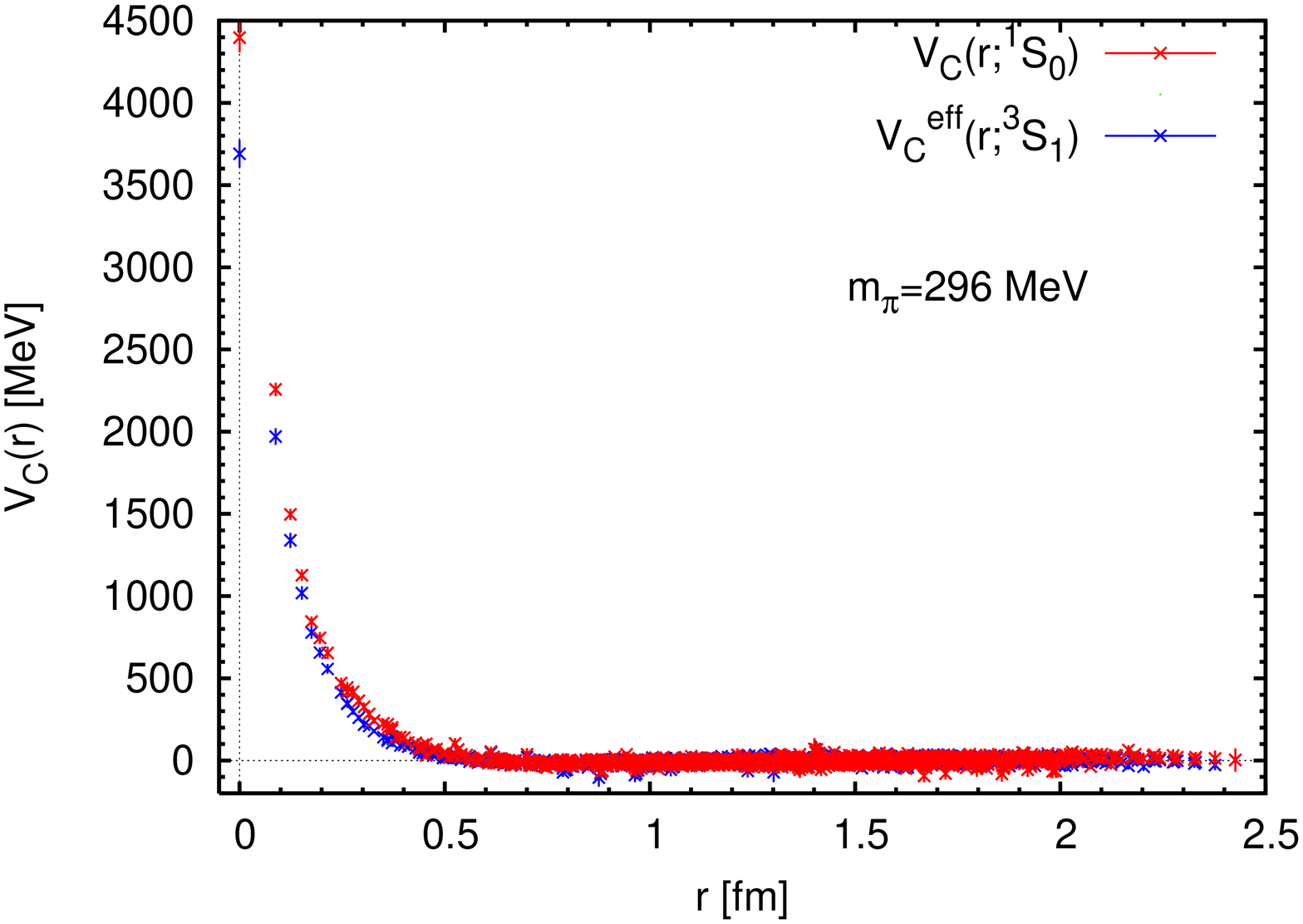}
\end{center}
\caption{Full  QCD  results of  (effective)  central potentials.   The
l.h.s            shows            the           results            for
$(\kappa_{ud},\kappa_{s})=(0.13700,0.13640)$,  and  the r.h.s.   shows
the results  for $(\kappa_{ud},\kappa_s)=(0.13770,0.13640)$, where the
inset is suppressed because of the huge error bar.}
\label{full.qcd.potentials}
\end{figure}
\Fig{full.qcd.potentials}(left)  shows  the full  QCD  results of  the
central   force   $V_{\rm    C}(r)\equiv   (E   -   H_0)\psi(\vec   x;
^1S_0)/\psi(\vec  x; ^1S_0)$  for  $^1S_0$ channel  and the  effective
central force $V_{\rm C}^{\rm  eff}(r; ^3S_1)\equiv (E - H_0)\psi(\vec
x;   ^3S_1)/\psi(\vec    x;   ^3S_1)$   for    $^3S_1$   channel   for
$(\kappa_{ud},\kappa_s)=(0.13700,0.13640)$.
Here,  $\psi(\vec x;^3S_1)$  denotes $\psi(\vec  x;^3S_1)\equiv [{\cal
P}\psi](\vec x)$ for notational simplicity.
$V_{\rm C}(r; ^1S_0)$ and $V_{\rm C}^{\rm eff}(r; ^3S_1)$ are obtained
from  BS   wave  functions  on   the  time-slices  $t=8$   and  $t=9$,
respectively, where  the ground state saturations  are achieved within
error bars.
Similar  to the quenched  results\cite{ishii,ishii.lattice.2007}, they
have attractive pockets of about  30 MeV in the medium distance, i.e.,
$0.5 \alt r \alt 1.0$ fm.
In contrast, the repulsive cores are considerably strong.
They  are by  about 10  times  as strong  as the  quenched result  for
comparable pion mass \cite{ishii.lattice.2007}.
There  seem to  be several  possible  reasons. (i)  A dynamical  quark
effect, (ii)  The action  adopted in quenched  calculation may  not be
close to the continuum limit.

\Fig{full.qcd.potentials}(right)  shows the  full QCD  results  of the
(effective)              central             potential             for
$(\kappa_{ud},\kappa_s)=(0.13770,0.13640)$. These results are obtained
from the BS  wave functions on the time-slice  $t=6$, where the ground
state saturations  are achieved within the statistical  errors.  We do
not show the inset because  of huge statistical errors. Although it is
necessary to  improve the statistics significantly to  reduce the huge
error bars,  we see  that the  repulsive cores  are  again considerably
strong.

\section{Summary}
We have presented preliminary lattice QCD results for the tensor force
by using  quenched QCD.
We have seen  that the tensor force has a  large quark mass dependence
and is enhanced as the quark mass decreases.
We have presented preliminary results of the (effective) central force
from 2+1 flavor lattice QCD  by using PACS-CS gauge configurations.  A
remarkable  difference from  the  quenched results  was  found in  the
strength of the repulsive core.
It is interesting  to investigate the reason, since  it may provide us
with a key  to the origin of  the repulsive core, which is  one of the
most important open problems in the nuclear physics.

\section*{Acknowledgments}
Quenched QCD  Monte Carlo  calculations have been  done with  IBM Blue
Gene/L at  KEK under a support  of its Large  Scale simulation Program
Nos. 06-21, 07-07, 08-19.
The  full QCD  calculations  have  been done  with  PACS-CS under  the
``Interdisciplinary  Computational  Science  Program'' of  Center  for
Computational Sciences, University of Tsukuba (No 08a-12).
We   thank   PACS-CS   Collaboration   for  the   2+1   flavor   gauge
configurations.
We are grateful for  authors and maintainers of {\it CPS++}\cite{cps},
of which a modified version is used for measurement done in this work.
N.I. thanks Dr.~T.~Izubuchi for a sample code of 3D FFT.
This  work was  supported  in  part by  Grant-in-Aid  of the  Japanese
Ministry of  Education, Science,  Support and Culture  (Nos. 18540253,
19540261, 20340047).

\end{document}